\documentclass [prb,twocolumn,showpacs,preprintnumbers,amsmath,amssymb]{revtex4}

\usepackage{graphicx}
\usepackage{dcolumn}

\newcommand{\beq}{\begin{equation}}
\newcommand{\eeq}{\end{equation}}
\newcommand{\beqa}{\begin{eqnarray}}
\newcommand{\eeqa}{\end{eqnarray}}

\begin{document}
\title{Relevance of phonon dynamics in strongly correlated systems coupled to phonons: 
\\ A Dynamical Mean Field Theory analysis}

\author{Giorgio Sangiovanni$^1$, Massimo Capone$^{2,3}$ and Claudio Castellani$^2$}

\affiliation{$^1$ Max-Planck Institut f\"ur Festk\"orperforschung, 
Heisenbergstr. 1, D-70569 Stuttgart, Germany \\
$^2$Istituto dei Sistemi Complessi, Consiglio Nazionale delle Ricerche,
Via dei Taurini 19, I-00185 Roma, Italy\\
$^3$ CNR-INFM Statistical Mechanics and Complexity Center, and Dipartimento di Fisica\\
Universit\`a di Roma "La Sapienza" piazzale Aldo Moro 5, I-00185 Roma, Italy}

\begin{abstract}
The properties of the electron-phonon interaction in the presence of a sizable 
electronic repulsion at finite doping are studied by investigating the 
metallic phase of the Hubbard-Holstein model with Dynamical Mean Field Theory. 
Analyzing the quasiparticle weight at finite doping, we find that a large 
Coulomb 
repulsion reduces the effect of  electron-phonon coupling at low-energy,
while this
reduction is not present at high energy.
The renormalization of the electron-phonon coupling induced by the Hubbard repulsion 
depends in a surprisingly strong and non-trivial way on the phonon frequency.
Our results suggest that phonon might affect differently high-energy and 
low-energy properties and this, together with the effect of phonon dynamics, should
be carefully taken into account when the effects of the electron-phonon interaction in a strongly 
correlated system, like the superconducting cuprates, are discussed.
\end{abstract}
\date{\today}
\maketitle

\section{Introduction} \label{sec:intro}

One of the open problems of high-temperature superconducting 
cuprates is the role played by the electron-phonon interaction.
The experimental evidences suggest a fairly strong influence 
of the electron-phonon interaction on some physical properties,
while for some other aspects, lattice effects seem to have almost 
no role. On one side, inelastic scattering measurements have shown 
that a specific optical in-plane phonon mode displays an anomalously 
pronounced softening.\cite{pint2,pint3,mcqueeney1999,dastuto}
The coupling to the same phonon mode has been invoked to explain the
kink  in the nodal electron dispersion detected by 
photoemission,\cite{kink1,lanzara1,kink2} 
On the other hand, the almost perfectly linear behavior of the resistivity
in a wide range of temperatures seems to indicate a little influence of 
phonons on transport properties. Isotope effects also show a complex phenomenology:
While the the superconducting temperature has a little isotope effect at optimal 
doping,\cite{isotope1,isotope2,isotope3} the in-plane penetration depth is much more 
sensitive to isotope substitution.\cite{isotope7,isotope4,isotope5}
Such a large (around 5\%) effect is usually translated into an isotope effect on the
effective mass suggesting the presence of polaronic carriers in underdoped compounds.\cite{isotope6}

The above puzzling scenario leads to a lively debate which ultimately focuses on 
 whether the effects of electron-phonon (e-ph) coupling on different 
quantities are depressed or enhanced by the presence of strong correlations.
Given the intrinsic nonperturbative character of the problem, there is no obvious
theoretical approach. Different approaches seem indeed to draw conflicting 
scenarios in studies of the Holstein model for the e-ph coupling
in the presence of strong correlations, described by the Hubbard 
or the $t$-$J$ models.
According to  Quantum Monte Carlo (QMC) and exact diagonalization (ED) 
a single hole in the $t$-$J$ model experiences
an enhanced polaronic effect due to the ``pre-localizing''
mechanism associated to the antiferromagnetic spin background.\cite{fehske-tJ,epj_io,
mishchenko,olle2}
QMC calculations in the Hubbard model at finite density and fairly high temperature
suggest that strong correlations favor the small transfered momentum electron-phonon 
vertex (or depress it less than its large momentum counterpart), 
and that this quantity increases by increasing repulsion for a window
of parameters.\cite{scalapino}
Slave boson approaches suggest however that such an effect is just a finite-temperature
precursor of a phase separation that would take place at low temperatures.\cite{bftPS,zeyher2}
Such tendency towards phase separation, indicated by mean-field approaches also 
for the three-band Hubbard model,\cite{PRB94} has been recently confirmed by the more accurate
Dynamical Mean Field Theory (DMFT).\cite{StJ1}

Another piece of information comes from the DMFT of the half-filled 
Hubbard-Holstein model in the paramagnetic sector (i.e., neglecting the 
antiferromagnetic ordering), 
where detailed phase diagrams are available.\cite{jeon,koller1} 
In Ref. \onlinecite{StJ2} we have 
shown that, close to the Mott transition, phonons have different effects on  
high- and low-energy single particle properties (self-energy, spectral weights,
density of states, \ldots). Namely, high-energy features are significantly affected by
phonons, whereas the low-energy quasiparticle features are basically
untouched and they coincide with those of an effective purely electronic 
with a slightly weaker repulsion $U$.
 In other words, the strong correlations strongly reduce the 
impact of e-ph interaction on quasiparticle properties, at least when the 
e-ph coupling is not too large. The only residual effect is a phonon-induced screening 
of $U$ which moreover is found to vanish linearly with the phonon frequency in the 
adiabatic limit.

This variety of results is certainly due to the different 
physical regimes they refer to. In this work we start from our analysis of 
the half-filled Hubbard-Holstein model and relax the half-filling condition,
therefore putting ourselves away from the Mott transition,
in a regime closer to that of other approaches.
We find that strong correlations still significantly harm the e-ph interaction.
In our strongly correlated metal polaronic behavior only establishes
for e-ph coupling larger by at least a factor two than the 
corresponding values in the absence of strong correlations or for a single hole 
in an antiferromagnetic background.
Moreover, we find that the interplay of electron-electron and e-ph interaction makes the value 
of the phonon frequency quite relevant, particularly for its effect on the effective 
mass. Remarkably, phonon dynamics turns out to be more relevant close to half-filling.

The paper is organized as follows: in sec. \ref{sec:model} we compare the 
half-filled and finite doping cases of the Hubbard-Holstein model. 
In sec. \ref{sec:zeta} we analyze the renormalization of the quasiparticle 
weight as a function of the electron-phonon coupling and the phonon frequency. 
In sec. \ref{sec:screening} the behavior of the chemical potential is studied
and the conclusions are then drawn in sec. \ref{sec:concl}.

\section{DMFT of the Hubbard-Holstein model: Half-filling vs Finite doping} \label{sec:model}
The Hamiltonian of the Hubbard-Holstein model reads
\begin{eqnarray}
H = & -t\sum_{\langle i,j\rangle,\sigma}  c^{\dagger}_{i,\sigma} c_{j,\sigma} 
+ U \sum_i n_{i\uparrow} n_{i\downarrow} \nonumber\\
& - g\sum_i n_i(a_i +a^{\dagger}_i) + \omega_0 \sum_i a^{\dagger}_i a_i,
\label{hamiltonian}
\end{eqnarray}
where $c_{i,\sigma}$ ($c^{\dagger}_{i,\sigma}$) and $a_i$ ($a^{\dagger}_i$) are, 
respectively, destruction (creation) operators for fermions with spin $\sigma$ and 
for local vibrations of frequency $\omega_0$ on site $i$, $t$ is the hopping 
amplitude, $U$ is the local Hubbard repulsion and $g$ is an electron-phonon 
coupling constant.
In this work we will always consider an infinite coordination Bethe-lattice
with semicircular density of states of semibandwidth $D$.
$\lambda=2g^2/\omega_0D$ is the standard electron-phonon coupling and $\omega_0/D$
is the adiabatic ratio.    

We solve the model by means of Dynamical Mean-Field Theory (DMFT), which in recent 
years has emerged as one of the most reliable tools for the analysis of both correlated 
materials and electron-phonon interactions. 
The method maps the lattice model onto an effective  local theory which still retains 
full quantum dynamics, and it is therefore expected to be quite accurate for 
models with local interactions as (\ref{hamiltonian}). 
The mean-field correspondence between the local theory and the original model is
achieved by
imposing a self-consistency condition which contains the information about the
original lattice.\cite{DMFTreview} 
In practice, the local problem is described through an Anderson-Holstein impurity 
model,\cite{olle84,hewsonam} whose impurity Green's function has to be calculated and
used to generate a new impurity model through the self-consistency equation.
To solve the impurity model we use ED,\cite{caffarel} truncating the 
infinite phonon Hilbert space allowing up to $N_{max}$ phonon states 
(ranging from $20$ to $40$), and using up to $N_b=9$ sites in the conduction 
bath.\cite{notashift}

DMFT has been widely use to study the properties of the Hubbard model, and 
a clear framework for the Mott-Hubbard transition has been determined.
Without entering the details of these studies, we just recall the
 qualitative difference between the half-filled system,
where a metal-insulator occurs at $U = U_{c2} \simeq 3D$, 
and finite-doping systems that have metallic character regardless the value of $U$. 
Such a distinction makes the effects of phonons different in the two cases.

At half-filling, in the strongly correlated metallic phase for  $U$
smaller, but not far from $U_{c2}$,
the Hubbard model presents a clear separation of 
energy scales, with high-energy Hubbard bands well separated from the
low-energy quasiparticle peak. In this regime the charge fluctuations are frozen, 
therefore strongly harming the e-ph interaction, at least with a Holstein coupling.
As discussed in Ref. \onlinecite{StJ2}, the residual effect of phonons
 can be described as a partial screening 
of the static Hubbard repulsion, which is ruled by 
the parameter $\omega_0/U$, according to 
the expression $U_{eff} = U - \eta\lambda D$, with $\eta$ given by
\begin{equation}\label{eta}
\eta=\frac{2\omega_0/U}{1+2\omega_0/U}.
\end{equation}
Notice that the above expression for $U_{eff}$ correctly reproduces 
the antiadiabatic limit $\omega_0/D \to \infty$, where the e-ph interaction mediates
an instantaneous local attraction of strength $\lambda$. 
The description in terms of an effective Hubbard model with a suitably
rescaled repulsion $U_{eff}$ works surprisingly well in the proximity of the Mott transition: 
The full Hubbard-Holstein model and the effective purely electronic model
display an identical electronic spectrum at low energy and, quite remarkably, 
the same rescaling for $U_{eff}$ makes the position of the Hubbard bands basically coincide. 
The effects of the electron-phonon coupling close to the Mott transition are 
well visible in the high-energy parts of the electronic spectrum, where phonon 
satellites show up, while the low-energy metallic peak turns out to be ``protected'' 
by correlation.

This paper is devoted to the extension of this analysis to generic fillings,
still limiting ourselves to the paramagnetic sector.
Conceptually the peculiarity of half-filling is the clear 
separation between high- and low-energy scales, which is rapidly lost as the
density deviates from one. This lack of hierarchy of energy scales makes 
it more complicated to draw a simple physical picture such as the one described above.
On the other hand, the properties of the metallic phase at finite doping are not 
severely affected by the antiferromagnetic order, which, if included, would instead
change considerably the half-filling picture. 

\section{Quasiparticle Properties} \label{sec:zeta}
As in   Ref. \onlinecite{StJ2}, we start our analysis from the
inspection of the quasiparticle weight
\begin{equation} \label{zeta}
Z = \left( 1-\left.\frac{\partial \Sigma'(\omega)}{\partial\omega} 
\right|_{\omega=0} \right)^{-1}
\end{equation} 
where $\Sigma'$ is the real part of the local self-energy $\Sigma$.
Due to the momentum independence of the self-energy in DMFT, $Z$ is also inversely 
proportional to the quasiparticle effective mass ratio $m^*/m$.
It is therefore clear that this quantity measures the metallic nature of the 
system, small values of $Z$ implying poorly metallic situations.

\subsection{Dependence on electron-phonon coupling} \label{subsec:lambda}
At half-filling, in the correlated metal for $U \lesssim U_{c2}$, the 
quasiparticle weight $Z$ \emph{increases} with the electron-phonon coupling 
$\lambda$.\cite{koller1,StJ2}
This counterintuitive behavior, in stark contrast with weakly correlated
systems,  where $Z$ decreases with $\lambda$,
is understood in terms of a phonon-driven attraction which counteracts the Hubbard
repulsion, and it can be quantitatively described
 within the effective picture introduced
in sec. \ref{sec:model}:\cite{StJ2}
By increasing $\lambda$, the effective repulsion decreases, thus 
 the system becomes less correlated, and the value of $Z$ increases.
As mentioned above, such a reduced effectiveness of the e-ph coupling
 can be associated to the freezing of  charge fluctuations, to which
the phonons are coupled. 
Close to Mott state, most sites are singly occupied, and 
doubly occupied and empty sites are a minority.
As soon as we move away from half-filling, doping, e.g., with holes, the number
of empty sites proliferates.
Therefore charge fluctuations are gradually restored, even if they are still 
reduced with respect to a system without electron correlations.
\begin{figure}[htbp]
\begin{center}
\includegraphics[width=8cm,height=7cm]{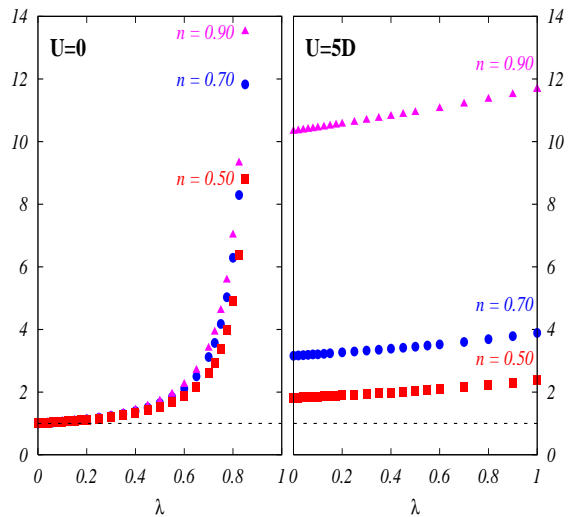}
\end{center}
\caption{(Color online) 
Calculated values of $m^*/m$ as a function of $\lambda$ for various values of the density. 
In both cases of $U=0$ (left) and $U=5D$ (right) the phonon frequency $\omega_0$ has been fixed to $0.2D$.}
\label{fig:U5vsU0}
\end{figure}
As a consequence, one can 
expect a stronger electron-phonon signature with respect to half-filling.
The behavior of $Z$ rapidly becomes more ordinary, namely $Z$ decreases as a 
function of $\lambda$ (except, as we will see below, for extremely large $\omega_0$ 
of the order of $U$) meaning that the predominant effect of the coupling to 
the lattice is of localizing nature.

This is also consistent with a much weaker dependence of $Z$ on $U$ for $U \simeq 5D$ 
(in comparison with the case $U \simeq U_c$ at half-filling) which
would depress the delocalizing effect of any phonon-induced variation of
$U$ into $U_{eff}$.

In Fig. \ref{fig:U5vsU0} we plot $m^*/m$ as a function of $\lambda$ for different doping levels and 
for $\omega_0=0.2D$. 
In the left panel we show the uncorrelated system ($U=0$).
 $m^*/m$ is equal to $1$ for $\lambda=0$, it rapidly increases with $\lambda$, 
eventually reaching a polaronic regime, testified by an exponential growth of
the effective mass and by the development of finite lattice distortions coupled 
to the electrons.\cite{massimo4} 
The right-hand panel presents instead a strongly correlated case($U=5D$).
Here the value of  $m^*/m$ at $\lambda=0$ is strongly dependent on density, since
correlations are more and more effective in localizing the carriers the closer we 
are to half-filling. In practice $m^*/m \simeq 2$ already for $n=0.50$ and it reaches
more than 10 for $n=0.9$.
The main result is however that the e-ph interaction is not able to substantially modify this values
determined by correlation, up to $\lambda = 1.0$-$1.5$ and $m^*/m$ is almost flat 
in this interval, if compared with the left panel. 
Thus, even if doped system do not show the growth of $Z$ with $\lambda$ characteristic
of half-filling (close to $U_{c2}$), the effect of e-ph coupling is anyway quantitatively reduced
by a sizable amount.
A similar information is brought by the location of  the polaron crossover:
\begin{figure}[htbp]
\begin{center}
\includegraphics[width=7.5cm]{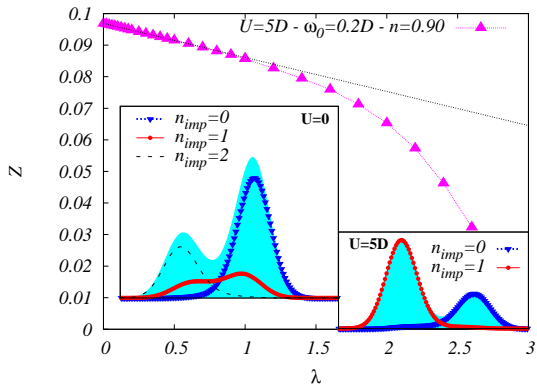}
\end{center}
\caption{(Color online) Behavior of the quasiparticle weight $Z$ (main panel) in the 
crossover from weak to strong $\lambda$ for $U=5D$, $n=0.90$ and $\omega_0=0.2D$. 
In the two insets the phonon distribution function $P(X)$ in the polaron region are 
shown for $U=5D$ (smaller inset, $\lambda=2.8$) and $U=0$ (larger one, $\lambda=0.875$). In both insets $n=0.70$ and $\omega_0=0.2D$.
The light blue shaded area is the full $P(X)$ while the triangles, solid circles and dashes represent $P(X)$ 
restricted to $n_{imp}=0$, $1$ and $2$ respectively.}
\label{fig:zeta_pol}
\end{figure}
If we consider larger value of $\lambda$ than in Fig. \ref{fig:U5vsU0}, the e-ph interaction
finally becomes able to sizably affect the quasiparticle residue $Z$. 
In the example shown in Fig. \ref{fig:zeta_pol} ($U=5D$, $n=0.9$, $\omega_0/D = 0.2$), 
the curve of $Z$ as a function of $\lambda$ clearly displays a crossover between a small-$\lambda$
linear behavior and a much faster decrease for  $\lambda \gtrsim 2\div 2.5$.
This is precisely the signature of a polaron crossover, as shown by many studies for the pure Holstein
model, where a similar bending of the curve occurs for much smaller  $\lambda \simeq 0.8$.\cite{massimo4,notahewson}

An alternative marker of the polaron crossover is the phonon displacement 
distribution $P(X)=\langle\psi_0\vert X\rangle\langle X\vert\psi_0\rangle$, 
where $\vert\psi_0\rangle$ is the groundstate vector, and $\vert X\rangle \langle X\vert$
is the projection operator on the subspace where the phonon displacement value $\hat{X} = 1/\sqrt{2M\omega_0}(a+a^{\dagger})$
($M$ being the phonon mass) has a given value $X$. This quantity therefore measures the distribution of the local distortions.
At weak coupling, $P(X)$  has only one peak, which only broadens when the coupling increases, but 
when a polaronic groundstate is realized, it presents two peaks, corresponding to different distortions
associated to the different charge states.\cite{notapol}
The two insets of Fig. \ref{fig:zeta_pol} show $P(X)$ for two sets of parameter
chosen in order to highlight the different way to realize  the bimodality  in 
the uncorrelated system (left/larger panel) and in the strongly correlated one
(right/small panel). 
The simple $P(X)$, shown as a shaded area, does not allow us to distinguish the
two situations.
Therefore we also plot the conditioned probability distributions
projected onto the states in which the impurity has different occupations
$n_{imp} = 0,1,2$.\cite{olle84}
For $U=0$ the bimodality is associated to a large number of empty and doubly
occupied sites (that gain more e-ph energy), and a smaller number of singly 
occupied ones. This is a signature of bipolaronic groundstate, where pairs of 
polarons are formed. Since we are at finite doping the two distributions have 
different shapes and heights. 
For $U=5D$ the two peaks of $P(X)$ are instead associated to empty sites 
and singly occupied ones, since the strong on-site repulsion unfavors
double occupancies. Exploiting the empty sites, the system can acquire 
polaronic groundstate at intermediate values of $\lambda$, while at half-filling 
this would require to completely overcome the Hubbard $U$.
We finally notice that in the special half-filling case $P(X)$ is not 
the ideal quantity to identify the polaronic behavior, which, 
 at intermediate values of $\lambda$, can only appear in 
the dynamics of the excitations (e.g. the behavior of a single hole)

To summarize the results of this section, we can conclude that for metallic situations
with finite doping it is not possible to describe the effects of the phonons on
quasiparticle properties in terms of an effective screened Coulomb repulsion, as it
happened at half-filling. 
It is anyway still true that the effect of phonons on quasiparticle properties
is substantially weaker than for weakly correlated systems, and that the phonon 
effects can be strong only in the high-energy part of the spectra.\cite{notaDOS}
The polaron crossover is found at values of $\lambda$ which are sensibly larger than
in the absence of strong correlation.

\subsection{Dependence on phonon frequency} \label{subsec:omega0}

As it has been discussed in many previous studies, the coupling
$\lambda$ is not the only parameter which controls the properties
of e-ph interaction, since also the phonon frequency plays an 
important role.\cite{massimo7,depolarone,epl} In this section we investigate precisely the role
of this quantity in strongly correlated systems.

\begin{figure}[htbp]
\begin{center}
\includegraphics[width=7.5cm]{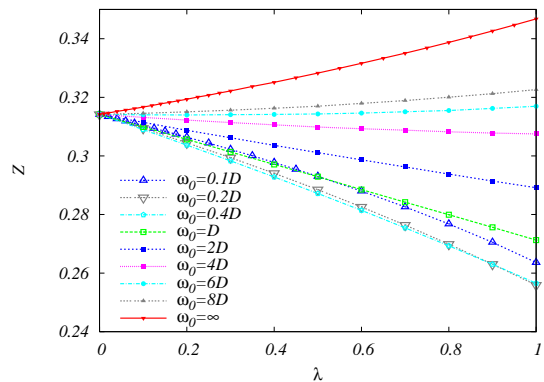}
\end{center}
\caption{(Color online) Quasiparticle weight $Z$ as a function of $\lambda$ for 
different values of $\omega_0/D$. $U=5D$ and $n=0.70$.}
\label{fig:zeta_n0.70}
\end{figure}

\begin{figure}[htbp]
\begin{center}
\includegraphics[width=7.5cm]{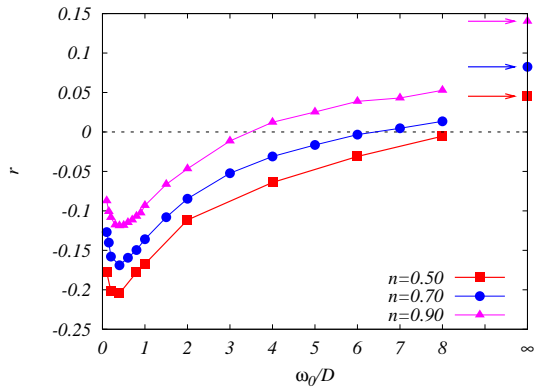}
\end{center}
\caption{(Color online) Coefficient $r$, defined as the slope of $Z/Z_{\lambda=0}$ for small $\lambda$ as a function of $\omega_0/D$, for $n=0.50$, $0.70$ and $0.90$.}
\label{fig:zcoeff}
\end{figure}

In Fig. \ref{fig:zeta_n0.70} we report $Z$ as a function 
of $\lambda$, for $U=5D$, $n=0.70$ and different values of $\omega_0$. 
The antiadiabatic curve (denoted by $\omega_0=\infty$) is simply obtained
for a  Hubbard model with total repulsion given by $U -\lambda D$. 
In the region $\lambda \lesssim 1$, well before any polaronic behavior, $Z$ 
is almost linear in $\lambda$, with a  slope strongly dependent on $\omega_0$. 
In Fig.\ref{fig:zcoeff} we plot the slope $r$, defined through $Z/Z_{\lambda=0}=1 + r  \lambda$.
For $\omega_0 \to \infty$, where the phonons only give rise to an instantaneous 
attraction, which opposes the Hubbard repulsion, $Z$ increases as a function of $\lambda$,
and $r > 0$. This behavior is indeed limited to $\omega_0 > 5D$, while for smaller
frequencies the phonons play a more standard role, decreasing $Z$, as expressed by
a negative $r$.
More interestingly, $r$ displays a non-monotonic behavior by further decreasing $\omega_0$.
Starting from $\omega_0=0$, it decreases up to $\omega_0 \simeq 0.3D$, roughly independently on doping, and then rises,
remaining negative for a wide range of frequencies, and eventually becoming positive for
quite large $\omega_0/D$.
While the evolution of $r$ from negative to positive can be simply understood in terms
of a crossover from an adiabatic region, where $\lambda$ acts as a localizing force,
to an antiadiabatic one where $\lambda$ decreases the localizing power of $U$,
the non-monotonic behavior suggests that the strongly correlated systems displays
more relevant energy scales.
\begin{figure}[htbp]
\begin{center}
\includegraphics[width=7.5cm]{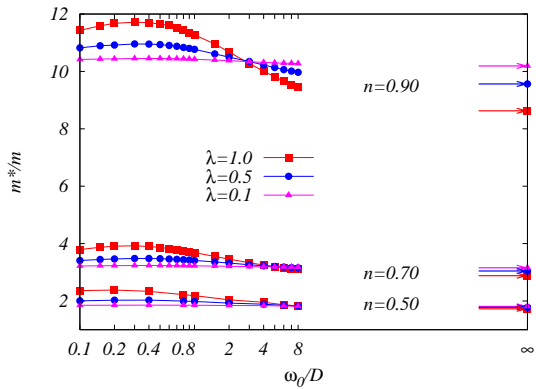}
\end{center}
\caption{(Color online) Quasiparticle effective mass vs. $\omega_0$ for $\lambda=0.1$, 
$0.5$ and $1.0$, at $U=5D$ in a semi-logarithmic plot. The three cases $n=0.50$, $0.70$
and $0.90$ are shown on the same scale.}
\label{fig:mstar}
\end{figure}
The same phenomenon can be observed by plotting the effective mass $m^*/m$ as 
a function of $\omega_0$ at fixed $\lambda$, as we do in Fig. \ref{fig:mstar} for 
the same three values of the doping. The derivative of this curve is
a measure of the isotope coefficient on the effective mass $\alpha_{m^*}$.
For pure e-ph systems $m^*$ always decreases with $\omega_0$ at fixed 
$\lambda$: In the adiabatic regime it initially decreases linearly with $\omega_0$ 
while, for $\omega_0 \gg D$, it scales as $1/\omega_0$.\cite{massimo9}
Therefore the isotope coefficient $\alpha_{m^*} = \frac{1}{2} \frac{\partial \log (m^*/m)}{\partial \log (\omega_0/D)}$ 
is always negative.
We find instead that, in the presence of strong correlation, 
$m^*$ can increase as a function of $\omega_0$.
This effect is larger for densities close to the density-driven 
metal-insulator transition: For instance, for $n=0.90$, the value of $\alpha_{m^*}$ 
for $\omega_0 \simeq 0.1D$ and $\lambda \simeq 1.0$ is a positive, even if quite small,
number $\alpha_{m^*} \simeq 1\div2 \times 10^{-2}$.
For values of $\omega_0$ around $0.3D$ the effective mass becomes instead almost 
independent on the phonon frequency, then $\alpha_{m^*}$ changes sign and $m^*/m$ 
monotonically approaches the asymptotic antiadiabatic value.
These results 
have an interesting consequence on the interpretation of the experiments on strongly
correlated materials: 
The sign and the magnitude of $\alpha_{m^*}$ is very sensitive to the values of the 
electron-phonon coupling and of the phonon frequency of the specific compound.\cite{massimo9,paola2}
Furthermore, a small value of $\alpha_{m^*}$ can be observed if a compound is
in the crossover region between the two regimes, but this cannot 
be interpreted as a sign of applicability of the Migdal-Eliashberg theory, 
despite this latter predicts indeed $\alpha_{m^*}=0$.
Turning to the cuprates, the sizable 
($\alpha_{m^*} \sim - 0.5$) negative isotopic effect for the effective mass
extracted  from the isotopic dependence of the penetration depth,\cite{isotope6} would suggest 
quite larger values of $\lambda$ than those studied here, 
putting these materials near (or beyond) the polaronic instability. 
It is well possible  that the antiferromagnetic correlations, neglected in this paper,
favor a polaronic behavior at least by decreasing doping.
Indeed the experimental evidences 
for a polaronic behavior in the cuprates are limited to extremely
small doping, well inside the antiferromagnetic phase.\cite{notacoll}

The isotope effect has instead a different behavior at half-filling for $U \lesssim U_{c2}$.
Close to the Mott-Hubbard transition $m^*$ decreases with $\omega_0$, but the 
absolute value of $\alpha_{m^*}$ is extremely large.\cite{StJ2,tesi}
This can be explained once more by the effective-Hubbard picture with
$U_{eff} = U-\eta \lambda D$ and $\eta = 2 \omega_0/U$ valid at half-filling and for 
$U \lesssim U_{c2}$:
close to the Mott-Hubbard transition, in fact, DMFT predicts that, for a pure Hubbard 
model with $U=U_{eff}$, $m^* \propto U_{c2} / (U_{c2} - U_{eff})$ i.e. 
\begin{equation}\label{m_vs_om}
m^* \propto \frac{U_{c2}}{U_{c2} - U + 2 \omega_0 \lambda D / U}
\end{equation}
An increase in the phonon frequency $\omega_0$ at fixed $\lambda$ determines then a 
sizable decrease in the effective mass.
The final result is $\alpha_{m^*} \propto -1/Z$, in other words the 
absolute value of $\alpha_{m^*}$ gets larger and larger, the closer one gets to the 
Mott transition.

 Finally we notice that the bare value of $r$ obtained in the absence of correlations
does not depend strongly on the phonon frequency up to $\omega_0 \sim D$. In the 
 adiabatic limit by $r_0=- N(E_F)$, where $N(E_F)$ is the non interacting 
density of states at the Fermi energy. For a semicircular density of states and
$n \simeq 0.5\div1$, one gets $r_0 \simeq - 0.58 \div 0.64$. Comparing with Fig. 4,
it turns out that this bare
value is larger by a factor $2 \div 6$ than the corresponding maximum values in the
interacting case.
If we define an effective coupling for the quasiparticles according to
 $1+r\lambda \equiv 1+r_0\lambda_{eff}$, we see that $\lambda_{eff}$ is rather smaller than 
$\lambda$ and this reduction is unusually enhanced at small frequency ($r_0$ will instead 
be maximum at $\omega_0=0$). At the same time, as discussed in sec. \ref{sec:zeta}, the polaronic
crossover is pushed to higher values of $\lambda$. 

\subsection{Fermi-liquid theory vs DMFT}\label{subsec:self-energy}

One of the main results of the previous analysis is that strong correlation
tends to  reduce the effects of e-ph interaction, and the size of this 
effect depends on the phonon frequency.
An important question is  whether or not the behavior of the
quasiparticle weight $Z$ obtained in DMFT in the metallic region away from  polaronic
instabilities can be captured, at least 
qualitatively, by a weak coupling approach to the interaction between quasiparticles 
and phonons.
In particular we can test  the validity of a Fermi-liquid analysis
in which a clear hierarchy is supposed: electron-electron correlations create
heavy quasiparticles, which in turn interact with the phonons via 
a renormalized density vertex. It is understood that DMFT is not able to deal
with the momentum dependence of the renormalized vertex.

We divide the self-energy in two contributions:

\begin{equation} \label{green1}
\Sigma({\bf k},\omega) =  \Sigma_{\lambda=0}({\bf k},\omega) + \Sigma_{res}({\bf k},\omega)
\end{equation}
where $\Sigma_{\lambda=0}$ is the self-energy of the pure Hubbard model, and 
$\Sigma_{res}$ contains all the additional interaction due to 
 bare and correlation-dressed electron-phonon processes.
We can define a quasiparticle
residue $Z_{\lambda=0}$ determined only by electronic correlations (We recall that
$\Sigma^{\prime}$ is the real part of $\Sigma$.)
\begin{equation} \label{zetaU}
Z_{\lambda=0}=\left( 1-\left.\frac{\partial \Sigma^{'}_{\lambda=0}({\bf k},\omega)}{\partial\omega} \right|_{\omega=0} \right)^{-1} 
\end{equation}
which is related to the full quasiparticle weight $Z$ by
\begin{equation} \label{zeta_qp}
\frac{Z}{Z_{\lambda=0}}=\left(1 - Z_{\lambda=0} \left.
\frac{\partial \Sigma^{'}_{res}({\bf k},\omega)}{\partial\omega} 
\right|_{\omega=0} \right)^{-1}.
\end{equation}
$Z/Z_{\lambda=0}$ may be seen as a wave-function renormalization for the quasiparticles
created by $U$  due to the additional interaction, and the quantity $Z_{\lambda=0}\Sigma'_{res}$ 
plays the role of a phonon-induced self-energy for those quasiparticles.

In the small-$\lambda$ regime, and assuming that the  quasiparticle self-energy 
is linear in $\omega$, 
Eq. (\ref{zeta_qp}) can be written as
\begin{equation}\label{zqp}
Z^{qp}=\frac{Z}{Z_{\lambda=0}}=1+r\lambda,
\end{equation}
where $r$ coincides with the phenomenological parameter introduced in sec. \ref{subsec:omega0} and plotted
in Fig. \ref{fig:zcoeff}.
Within Fermi-liquid theory we can derive an expression for $r$ at the lowest order in $\lambda$
assuming that $\Sigma_{res}$ only contains e-ph processes dressed by $U$: 
\begin{equation}\label{self-en1}
Z_{\lambda=0} \frac{\partial \Sigma'_{res}}{\partial \omega} =
 -N(E_F)^* V^* \Lambda^2 =  
- N(E_F) Z_{\lambda=0} \Lambda^2 \lambda D 
\end{equation}
where $N(E_F)^* = N(E_F)/Z_{\lambda=0}$ is the quasiparticle density of states ,
$V^* = \lambda D Z_{\lambda=0}^2$ is the renormalized e-ph interaction between 
quasiparticles, and $\Lambda$ is the vertex which couples the electrons
to the phonons (a density vertex for the Holstein model).
Eq. (\ref{self-en1}) obviously implies $r=-Z_{\lambda=0} \Lambda^2 N(E_F)$.
This result holds as long as one considers only those self-energy diagrams in 
which correlation dresses exclusively the vertex. 

There are only two specific limiting cases in which, making use of Ward 
Identities,\cite{nozieres,PRB94} we are able to find explicit expressions 
for the density vertex $\Lambda$: the static and the dynamic limit. 
These read respectively
\begin{equation}\label{Ward1}
Z_{\lambda=0} \Lambda (q \rightarrow 0,\omega=0) = \frac{1}{1+F_0^{s(e)}},
\end{equation}
and  
\begin{equation}\label{Ward2}
Z_{\lambda=0} \Lambda (q = 0,\omega \rightarrow 0) = 1
\end{equation}
where $F_0^{s(e)}$ is the symmetric Landau scattering amplitude due to electronic 
processes.
Since we are considering electronic processes only, we have, in the static limit
$\Lambda=\kappa^{(e)} / \kappa^{(e)}_0$
where $\kappa^{(e)}$ is the compressibility of the Fermi liquid in the absence of 
coupling to the lattice, while $\kappa^{(e)}_0$ is the non-interacting value 
equal to $2N(E_F)$.
Then Eq. (\ref{self-en1}) can be written, in the static limit, as
\begin{equation} \label{self-en3}
Z_{\lambda=0} \frac{\partial \Sigma'_{res}}{\partial \omega} 
= -  N(E_F) Z_{\lambda=0} \left[ \frac{\kappa^{(e)}}{\kappa_0^{(e)}} \right]^2 \lambda D 
\end{equation}
In the opposite dynamic limit, instead, $\Lambda=1 / Z_{\lambda=0}$ and 
Eq. (\ref{self-en1}) becomes
\begin{equation} \label{self-en5}
Z_{\lambda=0} \frac{\partial \Sigma'_{res}}{\partial \omega} 
= - \frac{N(E_F)}{Z_{\lambda=0}}  \lambda D \
\end{equation}

\begin{figure}[htbp]
\begin{center}
\includegraphics[width=7.5cm]{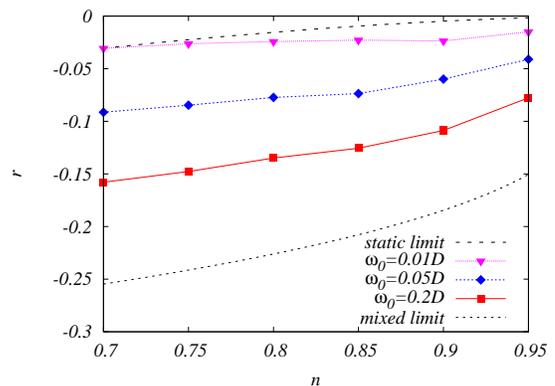}
\end{center}
\caption{Coefficient $r$ as a function of the density for three small values of $\omega_0/D$. The meaning of the curves labeled as static and mixed limit is described in the text.}
\label{fig:zcoeff2}
\end{figure}
We are now in the position to compare the above Fermi-liquid analysis with 
the DMFT results, where all the contributions to the electronic self-energy 
are considered.
In particular we focus on the small-$\omega_0$ regime, where the interplay
between attraction and repulsion is more subtle, as suggested by the data of Fig. 
\ref{fig:zcoeff}.
In Fig. \ref{fig:zcoeff2} we report the coefficient $r$, for $\omega_0/D=0.2$ and
for two smaller values and 
compare them, in the range of densities between $0.7$ and $0.95$ to the static 
limit result $r=-N(E_F) Z_{\lambda=0} ( \kappa^{(e)} / \kappa_0^{(e)} )^2$ of Eq. (\ref{self-en3}). 
The dynamic limit (\ref{self-en5}) is not shown because it gives $r \propto -1/Z_{\lambda=0}$, thus divergent 
in the limit $n\to 1$. This is evidently not comparable with the DMFT results shown
in Fig. \ref{fig:zcoeff2}, implying that this limit does not describe the numerical results.

The DMFT results are quite close to the static limit (\ref{self-en3}) only for the smallest
frequency we considered, but they rapidly move away as the frequency rises, and
already for $\omega_0 = 0.2D$ the difference between the calculation and the
Fermi-liquid prediction becomes huge. This means that a standard Fermi-liquid approach
can only be applied in the extremely adiabatic regime, and that even for small
phonon frequencies, there are quantitatively important corrections to the theory.

As an attempt  to provide a reference for the intermediate-$\omega_0$ regime, we 
considered a ``mixed limit'' solution, in which the vertices $\Lambda$ are 
taken as the geometric average between the two limiting expressions
(\ref{self-en3}) and (\ref{self-en5}). 
As one can see in Fig. \ref{fig:zcoeff2} this ``mixed limit'' solution has
a behavior similar to the  DMFT data. Interestingly, this heuristic choice
 has the advantage of staying finite for $n \to 1$, while the static limit goes to 
zero and appears to be appropriate only in the extremely small frequency limit.

\section{Screening of Coulomb repulsion by phonons} \label{sec:screening}

Differently from the half-filled case, in the regimes considered in
this paper we could not estimate $U_{eff}$ from the quasiparticle weight $Z$
since the delocalizing effect of the Coulomb screening turns out to be
irrelevant.
We can however discuss how the screening of Coulomb repulsion due to e-ph
interaction influences the behavior of the chemical potential $\mu$.

At half-filling $\mu$ is fixed by the particle-hole symmetry condition, 
while now, it needs to be determined in order to yield a given value of the 
density $n$.

\begin{figure}[htbp]
\begin{center}
\includegraphics[width=7.5cm]{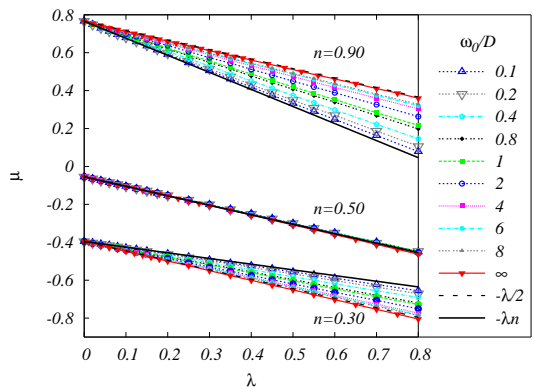}
\end{center}
\caption{(Color online) Behavior of the chemical potential as a function of $\lambda$ 
for different values of $\omega_0/D$. The cases of $n=0.90$, $n=0.50$ and $n=0.30$, all 
obtained for $U=5D$, are shown with the same scale. The full and dashed black lines are 
described below in the text.}
\label{fig:mu}
\end{figure}

In Fig. \ref{fig:mu} we report $\mu$ vs $\lambda$ as obtained within DMFT in the 
Hubbard-Holstein model varying the ratio $\omega_0/D$, comparing, for $U=5D$, 
the cases with $n=0.30$, $n=0.50$ and $n=0.90$.
For all densities $\mu$ decreases linearly in $\lambda$, while the role of $\omega_0$
changes with the value of the density:
for very small densities 
the slope of $\mu$ vs $\lambda$ monotonically decreases with increasing 
$\omega_0$ while it increases for $n$ close to $1$ and it is practically constant 
in $\omega_0$ at quarter filling ($n=0.50$).
Such a behavior of the chemical potential with $n$ can be explained 
taking a closer look to the self-energy $\Sigma(i\omega_n)$.
The real part of the self-energy at zero frequency shifts in fact the chemical 
potential in the interacting case according to $\mu = \mu_0 + \Sigma'(i\omega_n \to 0)$,
leading to $\mu_{\lambda} = \mu_{\lambda=0} + \Sigma'_{res}(i\omega_n \to 0)$.
A first indication on the behavior of $\Sigma(0)$ and $\mu$ can be derived 
through the Hartree-Fock (HF) approximation.
In this light it is useful to formally integrate out 
the phononic degrees of freedom in Eq. (\ref{hamiltonian}). 
The result is a retarded phonon-mediated interaction
\begin{equation}\label{Ueff}
V(\omega)= - \lambda D \frac{\omega_0^2}{\omega_0^2 -\omega^2}
\end{equation}
Including the Hubbard repulsion, the total interaction between electrons 
with opposite spin is then given by $U+2V(\omega)$, while only 
$V(\omega)$ is present between electrons with parallel spin. 
The Hartree term equals $(U-2\lambda D)\cdot n/2$. 
The interaction term $U+2V(\omega)$ between electrons with spin $\sigma$ and 
$-\sigma$ contributes in fact with $(U-\lambda D)\cdot n/2$ and the remaining 
$-\lambda D\cdot n/2$ comes from the term $V(\omega)$ between electrons with the 
same spin $\sigma$.

The equal-spin term gives also rise to a Fock diagram:
\begin{widetext}
\begin{equation}\label{Fock}
\Sigma^{Fock}_{\sigma}(i\omega_n) =
-\frac{\lambda D}{2} \int dE N(E) \left( 
\frac{\omega_0}{i\omega_n-E-\omega_0} \left[ f(E) + b(-\omega_0) \right] -
\frac{\omega_0}{i\omega_n-E+\omega_0} \left[ f(E) + b(\omega_0) \right] \right),
\end{equation}
\end{widetext}
where $f$ and $b$ denotes respectively the Fermi and Bose functions, 
and $N(E)$ is the density of state per spin.
In the small-$\omega_0$ limit this yields the well known $-\lambda D N(0) i\omega_n$
result,\cite{abrigorkov} and therefore the Hartree-Fock (HF) self-energy at zero
frequency is given by the Hartree term only, i.e.
\begin{equation}\label{SigmaHF_adiab}
\begin{array}{rl}
\Sigma^{HF}_{\sigma}(0) = \left(U-2\lambda D\right) \frac{n}{2} &
\hspace{1cm} (\omega_0 \rightarrow 0)
\end{array}
\end{equation}
In the opposite limit, i.e. for $\omega_0 \rightarrow \infty$ keeping $\lambda$ 
constant, it can be easily seen that Eq. (\ref{Fock}) is equal to 
$\lambda D \cdot (n-1)/2$, so that one obtains 
\begin{equation}\label{SigmaHF_antiadiab}
\begin{array}{rl}
\Sigma^{HF}_{\sigma}(0) = \left(U-\lambda D\right) \frac{n}{2} - 
\frac{\lambda D}{2} &
\hspace{1cm} (\omega_0 \rightarrow \infty)
\end{array}
\end{equation}

It is quite obvious that the HF scheme is not expected to hold for the
 large values of the Hubbard repulsion 
($U=5D$), we are dealing with.
For the pure Hubbard model it has indeed been found with iterated perturbation
theory solution of the DMFT, that the Hartree-Fock $U$-dependence can be canceled by
higher order contributions.\cite{kajueterPRL,potthoffIPT}
This finding is confirmed by ED-DMFT of the pure Hubbard model, where the low frequency self-energy is almost independent 
on $U$ for all $n<1$. \cite{notaSigma}

Therefore we can expect that  in the antiadiabatic limit, where the interaction is 
exactly a Hubbard one, the whole $U-\lambda D$
factor multiplying the density in the HF self-energy 
(see Eq. (\ref{SigmaHF_antiadiab})), is equally counteracted by higher-order terms.
This is confirmed by our results of Fig. \ref{fig:mu}, where 
$\mu(\lambda)-\mu_{\lambda=0} \simeq -\lambda D/2$ for all the considered densities in the antiadiabatic regime. 
This value is nothing but the constant term of Eq. (\ref{SigmaHF_antiadiab}),
which survives the cancellation of the density-dependent term. 

In the opposite adiabatic limit, we do not expect the phonon part of  (\ref{SigmaHF_adiab})
to follow the same fate of the purely electronic one. 
In this regime, in fact, the phonon-mediated attraction is 
of completely different nature with respect to the instantaneous Hubbard repulsion 
and the terms containing $\lambda$, beyond the HF, are not forced to  behave like 
the Hubbard term.
In other words, while for large $\omega_0$ the phonons induce a screening of the Coulomb
repulsion leading to an effective repulsion $U_{eff}\simeq U-\lambda D$, 
at vanishing phonon frequency this effect is not present and $U_{eff}\simeq U$.
On the basis of these considerations, in the limit $\omega_0/D \to 0$,
 we expect a cancellation only of the $U$-term in (\ref{SigmaHF_adiab}), leading to
$\mu(\lambda)-\mu_{\lambda=0} \simeq -\lambda D n$.
Such a relation is confirmed by the DMFT data for small $\omega_0$ and fits 
remarkably well with behavior obtained at all the considered values of the density, 
for the chemical potential (see the continuous line in Fig. \ref{fig:mu}).
This also clarifies the origin of the opposite behavior displayed by $\mu$ for 
$n < 1/2$ and $n > 1/2$.
In fact, as it can be seen still in Fig. \ref{fig:mu}, 
the fully antiadiabatic curve (denoted by $\omega_0=\infty$),\cite{notamu}
lays below all the other ones for $n=0.30$, contrary to the case of $n=0.90$ in 
which it lays above.
Moreover, still in Fig. \ref{fig:mu} it can be seen that the chemical potential is 
almost completely independent on $\omega_0$ for $n=0.50$, which is precisely what is 
predicted by the above considerations at quarter filling.
This supports our idea that the frequency dependence of
the $\lambda$-contribution to $\mu$ reflects the frequency dependence
of the Coulomb screening by phonons, even though we cannot put
this idea on a quantitative basis as we did instead 
near the Mott transition at half-filling.

\section{conclusions} \label{sec:concl}

In this paper we have studied the effects of the e-ph interaction 
on a strongly correlated metal for filling different from $n=1$.
Choosing a large value of $U/D$ the quasiparticle properties of the system are
mainly controlled by the electron-electron correlations and polaronic features
only appear at values of e-ph coupling $\lambda$ larger by at least a factor two 
than those in the absence of correlation. However, even if the effective 
e-ph coupling for quasiparticles
 is reduced by the large Hubbard repulsion, we have identified a strong
influence of the phonon dynamics. The quasiparticle weight $Z$, which in DMFT
is inversely proportional to the effective mass, becomes very small 
because of the localizing effect of the Hubbard repulsion but it is still 
substantially influenced by the value of the phonon frequency and displays an unusual isotope
effect.
Specifically, since $Z$ depends non monotonically on $\omega_0$, the isotope coefficient 
of the effective mass changes sign. This change of sign is a strong deviation with 
respect to  Migdal-Eliashberg theory, and it is not present in the half-filled system, 
except for peculiar situations close to the bipolaronic transition and for specific values
of $U/D$.\cite{paola2}
The effective quasiparticle e-ph coupling is particularly reduced in the small phonon
frequency limit. From this point of view our DMFT results bear similarities with a mean-field
calculation based on slave bosons and a variational Lang-Firsov transformation.\cite{paolo2}
In that approach one obtains near half-filling $Z/Z_{\lambda=0}\simeq 1- 1.388 (\omega_0/D) \lambda)$ at large $U/D$ and small $\omega_0/D$. In the adiabatic limit this expression implies a vanishingly small 
$\lambda_{eff}$, while in our DMFT $\lambda_{eff}$ remains finite for  $\omega_0\simeq 0$, 
as a result of the quantum fluctuations that are neglected in the mean-field calculation.
 These values of $\lambda_{eff}$ are however quite smaller than the bare $\lambda$, in 
agreement with a Fermi liquid description in terms of renormalized e-ph vertices in the static limit. 
It is interesting to note that the same mean-field approach/cite{paolo2} predicts at large $U/D$ a critical value 
$\lambda_{pol}\simeq U_c/2D$ for polaron formation which is slightly larger than the adiabatic value 
$\lambda_{pol}\simeq 1.328$
which is found for spinless fermions in the half-filled
Holstein model and somewhat smaller than our estimates $\lambda_{pol}\simeq 2\div 2.5 $ at $U=5D$.
\cite{notapaolo}
For a single hole in the $t$-$J$ with $J\simeq t/3$ 
a much smaller $\lambda_{pol}\simeq 0.8$
is required to reach a polaronic behavior. This is understood in terms of the pre-localizing mechanism
associated with the antiferromagnetic background.\cite{mishchenko,olle2,notaFL}

Beside the neglect of antiferromagnetic order and spin correlations,
our DMFT approach is limited by  the inability 
to capture the $k$-dependence of the self-energy, and accordingly of the 
quasiparticle weight.
This limitations does not allow us to describe the momentum dependence of the
effective e-ph coupling that has been proposed to be induced by strong correlations by various authors.\cite{scalapino,bftPS,zeyher2}
The momentum dependence can be restored by means of cluster extensions of
DMFT, such as the Cellular Dynamical Mean-Field Theory. It has already been
shown that, for the purely electronic Hubbard model, a significant momentum
dependence establishes when the Mott transition is approached already for
small clusters such as a $2 \times 2$ plaquette.\cite{massimo10}

A further crucial assumption of the present work is the Holstein form of the electron-phonon coupling 
term. Although this kind of description has been justified for the cuprates by explicit calculations
starting from a full three-band model,\cite{olle1} it must be noted that the symmetry of the 
electronic degrees of freedom which are coupled by phonons plays a very important 
role in the competition between the phonon-mediated attraction and the Coulomb 
repulsion. 
In a model for fullerenes, in fact, in which the electrons are mainly coupled 
to Jahn-Teller vibrations, it has been found that the attraction between electrons is not screened 
by a strong local repulsion, giving rise to unexpected phenomena like high-temperature 
superconductivity driven by strong correlation.\cite{massimo2,han1,massimo6}
More specifically in those models, phonon-mediated superconductivity is enhanced by the
strong correlations close to the Mott transition, due to the specific symmetry of the phonons,
which couple with orbital and spin degrees of freedom, as opposed to the Holstein coupling
with the electron density.

\section*{Acknowledgments}
We acknowledge useful discussions with Paolo Barone, Sergio Ciuchi, 
Michele Fabrizio, Erik Koch, 
Marco Grilli, Olle Gunnarsson, Paola Paci, Roberto Raimondi, and Alessandro Toschi.
This work has been financed by Miur PRIN Cofin 2003 and by CNR-INFM.



\begin{thebibliography}{54}
\bibitem[{not({\natexlab{a}})}]{notashift}
\bibinfo{note}{For large values of $\lambda$ it is particularly useful to work
  with shifted phonon variables, i.e. with a e-ph coupling term of the form $-g
  \sum_i (n_i - \bar{n}) (a_i + a^\dagger_i)$. This choice reduces the average
  number of phonons which otherwise increases linearly with $\lambda$.}
\bibitem[{not({\natexlab{c}})}]{notapol}
\bibinfo{note}{Notice that $P(X)$ is defined as a groundstate property while
  $Z$ is related to quasiparticle excitations so that in principle polaronic
  quasiparticles can also be present without a multimodal $P(X)$, as it indeed
  happens for a single hole in the $t$-$J$ model.}
\bibitem[{not({\natexlab{d}})}]{notaDOS}
\bibinfo{note}{In the spectral functions for large
  values of $U$, the shape of the metallic peak at the Fermi level is weakly
  affected by the e-ph interaction, while visible phonons satellites appear on
  top of the lower and upper Hubbard bands.}
\bibitem[{not({\natexlab{g}})}]{notaSigma}
\bibinfo{note}{While for $n<1$ and large $U$, both $\mu$ and $\mbox{Re}
  \Sigma_{\sigma}(0)$ are almost independent of $U$, the same two quantities
  are directly proportional to $U$ for $n>1$. In fact, the chemical potential
  of the metallic phase, in a Hubbard model with large $U$, is always close, or
  even within, to one of the two Hubbard bands and these are distant $U-2D$ one
  another.}
\bibitem[{not({\natexlab{h}})}]{notamu}
\bibinfo{note}{The antiadiabatic solution labeled by $\omega_0=\infty$ is
  calculated by solving a pure Hubbard model with total repulsion  $U
  -\lambda D$ and with chemical potential $\mu+\lambda D/2$. }
\bibitem[{not({\natexlab{i}})}]{notapaolo}
\bibinfo{note}{Notice that the mean-field estimate of Ref. \onlinecite{paolo2}
  increases by decreasing $U/D$ up to $\lambda_{pol}\simeq U_c/D$ near $U\simeq
  U_c$.}
\bibitem[{not({\natexlab{j}})}]{notaFL}
\bibinfo{note}{It is worth to note that also the FL description would suggest a
  greater relevance of e-ph coupling in the AF phase because of the larger
  value $Z\simeq J/t$ of the quasiparticle weight with respect to the
  paramagnetic phase at small doping.}
\end{thebibliography}
\end{document}